\documentclass[12pt]{article}
\usepackage{a4wide}
\usepackage{latexsym}
\usepackage{amsmath}
\usepackage{amsfonts}
\usepackage{amscd}
\usepackage{cite}

\usepackage{pslatex}
\usepackage{graphicx}
\usepackage[latin1]{inputenc}
\usepackage[T1]{fontenc}

\newcommand{\bq}{\begin{eqnarray}}
\newcommand{\eq}{\end{eqnarray}}

\begin{document}

\thispagestyle{empty}

\begin{flushright}
  MITP/14-037 \\
  MaPhy-AvH/2014-04
\end{flushright}

\vspace{1.5cm}

\begin{center}
  {\Large\bf The two-loop sunrise graph in two space-time dimensions with arbitrary masses
             in terms of elliptic dilogarithms\\
  }
  \vspace{1cm}
  {\large Luise Adams ${}^{a}$, Christian Bogner ${}^{b}$ and Stefan Weinzierl ${}^{a}$\\
  \vspace{1cm}
      {\small ${}^{a}$ \em PRISMA Cluster of Excellence, Institut f{\"u}r Physik, }\\
      {\small \em Johannes Gutenberg-Universit{\"a}t Mainz,}\\
      {\small \em D - 55099 Mainz, Germany}\\
  \vspace{2mm}
      {\small ${}^{b}$ \em Institut f{\"u}r Physik, Humboldt-Universit{\"a}t zu Berlin,}\\
      {\small \em D - 10099 Berlin, Germany}\\
  } 
\end{center}

\vspace{2cm}

% abstract ---------------------------------------
\begin{abstract}\noindent
  {
We present the two-loop sunrise integral with arbitrary non-zero masses in two space-time dimensions
in terms of elliptic dilogarithms.
We find that the structure of the result is as simple and elegant as in the equal mass case, 
only the arguments of the elliptic dilogarithms are modified.
These arguments have a nice geometric interpretation.
   }
\end{abstract}

\vspace*{\fill}

% -----------------------------------------------------------------------------
\newpage

\section{Introduction}
\label{sec:intro}

The study of transcendental functions appearing in the computation of Feynman loop integrals is a fascinating
subject.
At the one-loop level, there are just two transcendental functions appearing: the logarithm and the dilogarithm.
These generalise to multiple polylogarithms, observed in higher loop integrals.
However, there are Feynman integrals which cannot be expressed in terms of multiple polylogarithms.
The two-loop sunrise integral with non-zero masses is the simplest example of such an integral.
This is a non-trivial integral and has received considerable attention in the 
literature 
\cite{Broadhurst:1993mw,Berends:1993ee,Bauberger:1994nk,Bauberger:1994by,Bauberger:1994hx,Caffo:1998du,Laporta:2004rb,Groote:2005ay,Groote:2012pa,Bailey:2008ib,MullerStach:2011ru,Adams:2013nia,Bloch:2013tra,Remiddi:2013joa,Caffo:2002ch,Pozzorini:2005ff,Caffo:2008aw}.
The two-loop sunrise integral with non-zero masses is relevant 
for precision calculations in electro-weak physics, where non-zero masses naturally occur.
In addition, it appears as a sub-topology in many advanced higher-order calculations,
like the two-loop corrections to top-pair production or in the computation of higher-point functions in massless theories.
It is therefore desirable to understand this integral in detail, 
as such an understanding will pave the way for more complicated processes.

In this paper we discuss the two-loop sunrise integral 
with non-zero masses in two space-time dimensions.
The restriction to two space-time dimensions allows us to focus on the essential difficulties 
and avoids some entanglements with complications which we know how to handle.
In two space-time dimensions the integral is finite and depends only on the second graph polynomial,
but not on the first graph polynomial.
Working in two space-time dimensions removes therefore the issues of ultraviolet divergences
and the dependence on the first graph polynomial.
Of course we are interested in the end in the result around four space-time dimensions.
Using dimensional recurrence relations \cite{Tarasov:1996br,Tarasov:1997kx}, 
the result in two space-time dimensions can be related to the result in four space-time dimensions.
For the sunrise integral this has been discussed recently in detail in \cite{Remiddi:2013joa}.
However, the result around four space-time dimensions will require two additional integrals.
There is some freedom in the choice of these additional integrals, one possibility is given by
the integrals $I_3$ and $I_4$ of eq.~(51) in \cite{MullerStach:2011ru},
a second possibility is given by the integrals $Z_2^{(1)}$ and $Z_3^{(1)}$ of eq.~(4.11) in \cite{Remiddi:2013joa}.
These additional integrals are not subject of this article.
The result for the sunrise integral around four space-time dimensions will be addressed in a future publication.

It has been known for quite a while that in the equal mass case the sunrise integral 
satisfies an ordinary second-order linear differential equation \cite{Broadhurst:1993mw}.
The analytic solution up to quadrature in the equal mass case has been given in \cite{Laporta:2004rb}.
Renewed interest in the sunrise integral arose with the observation, 
that the two-loop sunrise integral with unequal masses in two space-time dimensions 
satisfies as well an ordinary second-order linear differential equation \cite{MullerStach:2011ru}. 
The corresponding analytic solution (up to quadrature) in the unequal mass case was presented in \cite{Adams:2013nia}.

Methods of algebraic geometry are very helpful in the study of the massive two-loop sunrise integral.
The two-loop sunrise integral is closely related to an elliptic curve.
This elliptic curve is defined by the (second) graph polynomial.
The order of the ordinary linear differential equation for the integral follows from the dimension of the first cohomology group
of the elliptic curve (which is two).
The two solutions of the homogeneous differential equation are periods of the elliptic curve.

Recently, it was shown in the equal mass case that the solution of 
the inhomogeneous differential equation can be written as a product of a period with an elliptic dilogarithm \cite{Bloch:2013tra}.
In this paper we consider the more general case of unequal non-zero masses
and we present the solution for the two-loop sunrise integral with arbitrary non-zero masses in two space-time dimensions
in terms of periods and elliptic dilogarithms.
To our surprise, the result is almost as simple as the corresponding result in the equal mass case,
only the arguments of the elliptic dilogarithms are modified.
These arguments have a nice geometric interpretation as the images of the intersection points of the variety defined by
the graph polynomial with the integration region under a chain of mappings, which we describe in detail.

This paper is organised as follows:
In the next section we define the two-loop sunrise integral in two space-time dimensions and introduce our notation.
In section~\ref{sec:differential_equation} we review the second-order differential equation. 
The boundary value at $t=0$ is shortly discussed in section~\ref{sec:boundary}.
In section~\ref{sec:hom_solutions} we present a streamlined derivation for the solutions of the homogeneous differential equation.
In section~\ref{sec:change_variables} we start the discussion of the solution for the inhomogeneous differential equation 
and first change variables by replacing the momentum squared with the nome.
In section~\ref{sec:equal} we briefly discuss the solution of the inhomogeneous differential equation in the equal mass case.
This motivates the generalisation of the classical polylogarithms towards the elliptic case, which is given in section~\ref{sec:polylogs}.
The solution for the two-loop sunrise integral in the unequal mass case in terms of elliptic dilogarithms 
is presented in section~\ref{sec:unequal_mass_case} and is the main result of this paper.
The geometric interpretation of the arguments of the elliptic dilogarithms is given in section~\ref{sec:geometric_interpretation}.
Finally, section~\ref{sec:conclusions} contains our conclusions.
Manipulations involving elliptic integrals, elliptic functions and related expressions are not (yet) standard in the field of Feynman loop integral
calculations.
For the convenience of the reader we included an appendix, containing details of our calculation.

% ------------------------------------------------------------------------------------------
\section{Definition of the two-loop sunrise integral and notation}
\label{sec:definition}

The two-loop sunrise integral with arbitrary masses is given in $2$-dimensional Minkowski space by
\bq
\label{def_sunrise}
 S\left( p^2 \right)
 & = &
 \mu^2
 \int \frac{d^2k_1}{i \pi} \frac{d^2k_2}{i \pi}
 \frac{1}{\left(-k_1^2+m_1^2\right)\left(-k_2^2+m_2^2\right)\left(-\left(p-k_1-k_2\right)^2+m_3^2\right)}.
\eq
The three internal masses are denoted by $m_1$, $m_2$ and $m_3$. 
The arbitrary scale $\mu$ is introduced to make the integral dimensionless.
$p^2$ denotes the momentum squared. 
In terms of Feynman parameters the two-loop integral is given by
\bq
\label{def_Feynman_integral}
 S\left( p^2\right)
 & = & 
 \mu^2
 \int\limits_\sigma \frac{\omega}{{\cal F}} 
\eq
with
\bq
 {\cal F} & = & - x_1 x_2 x_3 p^2
                + \left( x_1 m_1^2 + x_2 m_2^2 + x_3 m_3^2 \right) \left( x_1 x_2 + x_2 x_3 + x_3 x_1 \right).
\eq
The differential two-form $\omega$ is given by
\bq
 \omega & = & x_1 dx_2 \wedge dx_3 + x_2 dx_3 \wedge dx_1 + x_3 dx_1 \wedge dx_2.
\eq
Eq.~(\ref{def_Feynman_integral}) is a projective integral and we can integrate over any surface covering the solid angle of the first octant:
\bq
\label{def_sigma}
 \sigma & = & \left\{ \left[ x_1 : x_2 : x_3 \right] \in {\mathbb P}^2 | x_i \ge 0, i=1,2,3 \right\}.
\eq
It will be convenient to introduce the notation
\bq
 t & = & p^2
\eq
for the momentum squared.
We denote the masses related to the pseudo-thresholds by
\bq
\label{def_pseudo_thresholds}
 \mu_1 = m_1+m_2-m_3,
 \;\;\;
 \mu_2 = m_1-m_2+m_3,
 \;\;\;
 \mu_3 = -m_1+m_2+m_3,
\eq
and the mass related to the threshold by
\bq
\label{def_thresholds}
 \mu_4 = m_1+m_2+m_3.
\eq
We further set
\bq
\label{def_D}
 D & = & 
 \left( t - \mu_1^2 \right)
 \left( t - \mu_2^2 \right)
 \left( t - \mu_3^2 \right)
 \left( t - \mu_4^2 \right).
\eq
We also introduce the monomial symmetric polynomials $M_{\lambda_1 \lambda_2 \lambda_3}$ in the variables $m_1^2$, $m_2^2$ and $m_3^2$.
These are defined by
\bq
 M_{\lambda_1 \lambda_2 \lambda_3} & = &
 \sum\limits_{\sigma} \left( m_1^2 \right)^{\sigma\left(\lambda_1\right)} \left( m_2^2 \right)^{\sigma\left(\lambda_2\right)} \left( m_3^2 \right)^{\sigma\left(\lambda_3\right)},
\eq
where the sum is over all distinct permutations of $\left(\lambda_1,\lambda_2,\lambda_3\right)$.
In addition, we introduce the abbreviations
\bq
\label{def_delta}
% Delta is -Delta_3 of hep-ph/0502165
\Delta = \mu_1 \mu_2 \mu_3 \mu_4,
 \;\;\;\;\;\;
 \delta_1 = -m_1^2 + m_2^2 + m_3^2,
 \;\;\;\;\;\;
 \delta_2 = m_1^2 - m_2^2 + m_3^2,
 \;\;\;\;\;\;
 \delta_3 = m_1^2 + m_2^2 - m_3^2
\eq
and
\bq
\label{def_v_variables}
 v_1 = \frac{1+i\frac{\sqrt{\Delta}}{\delta_1}}{1-i\frac{\sqrt{\Delta}}{\delta_1}},
 \;\;\;\;\;\;
 v_2 = \frac{1+i\frac{\sqrt{\Delta}}{\delta_2}}{1-i\frac{\sqrt{\Delta}}{\delta_2}},
 \;\;\;\;\;\;
 v_3 = \frac{1+i\frac{\sqrt{\Delta}}{\delta_3}}{1-i\frac{\sqrt{\Delta}}{\delta_3}}.
\eq
We denote the complex conjugate of $v$ by $v^\ast$.
The variables $v_1$, $v_2$ and $v_3$ are complex numbers of unit norm 
and hence $v_i^{-1}=v_i^\ast$ for $i =1,2,3$.

Without loss of generality we assume that the masses are ordered as
\bq
\label{ordering_masses}
 0 < m_1 \le m_2 \le m_3.
\eq
We start our calculation in the Euclidean region, defined by
\bq
 -t & \ge & 0,
\eq
and in the vicinity of the equal mass point.
In the vicinity of the equal mass point we can assume $m_3 < m_1+m_2$, or equivalently $\mu_1>0$.
With this assumption and with the ordering of eq.~(\ref{ordering_masses}) we have
\bq
 0 < \mu_1 \le \mu_2 \le \mu_3 < \mu_4.
\eq
In addition we can assume in the vicinity of the equal mass point that
\bq
 \delta_i > 0, & & i \in \{1,2,3\}.
\eq
Our result can be continued analytically to all other regions of interest by
Feynman's $i0$-prescription, where we substitute $t \rightarrow t + i0$. 
The symbol $+i0$ denotes an infinitesimal positive imaginary part.

% ------------------------------------------------------------------------------------------
\section{The differential equation}
\label{sec:differential_equation}

The two-loop sunrise integral in two dimensions satisfies 
a second-order differential equation in the variable $t=p^2$ \cite{MullerStach:2011ru}:
\bq
\label{second_order_dgl}
 \left[ p_2(t) \frac{d^2}{d t^2} + p_1(t) \frac{d}{dt} + p_0(t) \right] S\left(t\right) & = & \mu^2 p_3(t),
\eq
where $p_2(t)$, $p_1(t)$, $p_0(t)$ and $p_3(t)$ are polynomials in $t$.
The polynomial $p_2(t)$, $p_1(t)$ and $p_0(t)$ appearing on the left hand side within the Picard-Fuchs operator are given by
\bq
\label{def_p0}
 p_2(t) & = &
  t \left( t - \mu_1^2 \right)
    \left( t - \mu_2^2 \right)
    \left( t - \mu_3^2 \right)
    \left( t - \mu_4^2 \right)
    \left( 3 t^2 - 2 M_{100} t + \Delta \right),
 \nonumber \\
 p_1(t) & = & 
  9 t^6
  - 32 M_{100} t^5
  + \left( 37 M_{200} + 70 M_{110} \right) t^4
  - \left( 8 M_{300} + 56 M_{210} + 144 M_{111} \right) t^3
 \\
 & &
  - \left( 13 M_{400} - 36 M_{310} + 46 M_{220} - 124 M_{211} \right) t^2
 \nonumber \\
 & &
  - \left( -8 M_{500} + 24 M_{410} - 16 M_{320} - 96 M_{311} + 144 M_{221} \right) t
 \nonumber \\
 & &
  - \left( M_{600} - 6 M_{510} + 15 M_{420} - 20 M_{330} + 18 M_{411} - 12 M_{321} - 6 M_{222} \right),
 \nonumber \\
 p_0(t) & = &
  3 t^5
  - 7 M_{100} t^4
  + \left( 2 M_{200} + 16 M_{110} \right) t^3
  + \left( 6 M_{300} - 14 M_{210} \right) t^2
 \nonumber \\
 & &
  - \left( 5 M_{400} - 8 M_{310} + 6 M_{220} - 8 M_{211} \right) t
  + \left( M_{500} - 3 M_{410} + 2 M_{320} + 8 M_{311} - 10 M_{221} \right).
 \nonumber
\eq
The polynomial $p_3(t)$ appearing in the inhomogeneous part is given by
% rational piece in factorised form
\bq
 p_3(t) & = & 
 -2 \left( 3 t^2 - 2 M_{100} t + \Delta \right)^2
 \\
 & & 
 + 2 c\left(t,m_1,m_2,m_3\right)  \ln \frac{m_1^2}{\mu^2}
 + 2 c\left(t,m_2,m_3,m_1\right)  \ln \frac{m_2^2}{\mu^2}
 + 2 c\left(t,m_3,m_1,m_2\right)  \ln \frac{m_3^2}{\mu^2},
 \nonumber 
\eq
with
\bq
\lefteqn{
c\left(t,m_1,m_2,m_3\right) = } & &
 \nonumber \\
 & &
 \left( -2 m_1^2 + m_2^2 + m_3^2 \right) t^3
 + \left( 6 m_1^4 - 3 m_2^4 - 3 m_3^4 - 7 m_1^2 m_2^2 - 7 m_1^2 m_3^2 + 14 m_2^2 m_3^2 \right) t^2
 \nonumber \\
 & &
 + \left( -6 m_1^6 + 3 m_2^6 + 3 m_3^6 + 11 m_1^4 m_2^2 + 11 m_1^4 m_3^2 - 8 m_1^2 m_2^4 - 8 m_1^2 m_3^4 - 3 m_2^4 m_3^2 - 3 m_2^2 m_3^4 \right) t
 \nonumber \\
 & & 
 + \left( 2 m_1^8 - m_2^8 - m_3^8 - 5 m_1^6 m_2^2 - 5 m_1^6 m_3^2 + m_1^2 m_2^6 + m_1^2 m_3^6 + 4 m_2^6 m_3^2 + 4 m_2^2 m_3^6 
 \right. \nonumber \\
 & & \left.
        + 3 m_1^4 m_2^4 + 3 m_1^4 m_3^4 - 6 m_2^4 m_3^4 
        + 2 m_1^4 m_2^2 m_3^2 - m_1^2 m_2^4 m_3^2 - m_1^2 m_2^2 m_3^4 \right).
\eq

%------------------------------------------------------------------------------
\section{The boundary value}
\label{sec:boundary}

We denote by $S_0$ the boundary value of the two-loop sunrise integral at $t=0$: $S_0=S(0)$.
In a neighbourhood of the equal mass point we have $\Delta>0$ and the integral $S_0$ is expressed in the region
$\Delta > 0$  by \cite{Ussyukina:1993jd,Lu:1992ny,Bern:1997ka,Adams:2013nia}
\bq
\label{boundary_1}
 S_0 = 
 \frac{2\mu^2}{\sqrt{\Delta}} 
 \left[ \mathrm{Cl}_2\left( \alpha_1 \right) + \mathrm{Cl}_2\left( \alpha_2 \right) + \mathrm{Cl}_2\left( \alpha_3 \right) \right].
\eq
The arguments $\alpha_i$ ($i \in \{1,2,3\}$) of the Clausen functions are given by
\bq
 \alpha_i & = & 2 \arctan \left( \frac{\sqrt{\Delta}}{\delta_i}\right).
\eq
The quantities $\delta_i$ and $\Delta$ have been defined in eq.~(\ref{def_delta}).
Note that we have
\bq
 \mathrm{Cl}_2\left(\alpha_i\right) & = & 
 \frac{1}{2i} \left[ \mathrm{Li}_2\left(v_i\right) - \mathrm{Li}_2\left(v_i^{-1}\right) \right],
\eq
where the variables $v_i$ have been defined in eq.~(\ref{def_v_variables}).
The variables $\alpha_i$ and $v_i$ are related by
\bq
 v_i & = & e^{i \alpha_i}.
\eq
We thus have
\bq
\label{boundary_2}
 S_0 = 
 \frac{\mu^2}{i \sqrt{\Delta}} 
 \left[ 
        \mathrm{Li}_2\left( v_1 \right) + \mathrm{Li}_2\left( v_2 \right) + \mathrm{Li}_2\left( v_3 \right) 
      - \mathrm{Li}_2\left( v_1^{-1} \right) - \mathrm{Li}_2\left( v_2^{-1} \right) - \mathrm{Li}_2\left( v_3^{-1} \right) 
 \right].
\eq
The form of the boundary integral in
eq.~(\ref{boundary_2}) will be slightly more convenient in the sequel.
In the equal mass case ($m_1=m_2=m_3=m$) eq.~(\ref{boundary_2}) reduces to
\bq
\label{boundary_equal_mass}
 S_0 = 
 \frac{3 \mu^2}{i \sqrt{\Delta}} 
 \left[ 
        \mathrm{Li}_2\left( r_3 \right) 
      - \mathrm{Li}_2\left( r_3^{-1} \right) 
 \right],
\eq
where $r_q$ denotes the $q$-th root of unity:
\bq
 r_q & = & \exp\left(\frac{2\pi i}{q}\right).
\eq

%------------------------------------------------------------------------------

\section{The homogeneous solutions}
\label{sec:hom_solutions}

In this section we present a streamlined derivation of the homogeneous solutions of the differential equation in eq.~(\ref{second_order_dgl}).
We note that the equation ${\cal F}=0$ has been used in \cite{MullerStach:2011ru} to determine the Picard-Fuchs operator in eq.~(\ref{second_order_dgl}).
In \cite{Adams:2013nia} we showed that the solutions of the homogeneous differential equation are the periods of an auxiliary elliptic curve.
This auxiliary elliptic curve was guessed from the equal mass limit and the imaginary part of the integral.
However, the appearance of the auxiliary elliptic curve seems ad-hoc and not quite satisfactory.
In this section we show that there is no need for an auxiliary elliptic curve. 
We show that the solutions of the homogeneous differential equation are the periods of the elliptic curve defined by
the zero set of the graph polynomial ${\mathcal F}$.
The equation
\bq
\label{eq_F_0}
 {\cal F} & = & 0
\eq
defines as it stands a cubic curve in ${\mathbb C}{\mathbb P}^2$.
With the choice of a rational point as the origin $O$, this curve becomes an elliptic curve.
Rational points are for example the three intersection points of the integration region $\sigma$ (defined in eq.~(\ref{def_sigma}))
with the variety defined by ${\mathcal F}=0$. These points are given by
\bq
\label{intersection_F_sigma}
 P_1 = \left[1:0:0\right], 
 \;\;\;
 P_2 = \left[0:1:0\right], 
 \;\;\;
 P_3 = \left[0:0:1\right].
\eq
We will choose one of these three points $P_1$, $P_2$, $P_3$ as the origin $O$.
We denote the corresponding elliptic curves by $E_1$, $E_2$ and $E_3$, where the index refers to the choice of the origin.
The elliptic curve $E_i$ (with $i \in \{1,2,3\}$) can be transformed into the Weierstrass normal form
\bq
\label{WNF_with_g2_g3}
 \hat{E} & : & y^2 z = 4 x^3 - g_2 x z^2 - g_3 z^3.
\eq
Under this change of variables the origin $O$ of $E_i$ is transformed to the point $[x:y:z]=[0:1:0]$,
which is the origin (or the point at infinity) of the elliptic curve $\hat{E}$.
The explicit expressions for this change of variables are given in appendix~\ref{trafo_to_WNF}.
Note that the same Weierstrass normal form is obtained for $E_1$, $E_2$ and $E_3$.
In the following we will work in the chart $z=1$.
Factorising the cubic polynomial on the right-hand side of eq.~(\ref{WNF_with_g2_g3}),
the Weierstrass normal form can equally be written as
\bq
 y^2 & = & 4 \left(x-e_1\right)\left(x-e_2\right)\left(x-e_3\right),
 \;\;\;\;\;\;
 \mbox{with} 
 \;\;\;
 e_1+e_2+e_3=0,
\eq
and
\bq
 g_2 = -4 \left( e_1 e_2 + e_2 e_3 + e_3 e_1 \right),
 & &
 g_3 = 4 e_1 e_2 e_3.
\eq
The roots are given by
\bq
\label{def_roots}
 e_1 & = & \frac{1}{24 \mu^4} \left( -t^2 + 2 M_{100} t + \Delta + 3 \sqrt{D} \right),
 \nonumber \\
 e_2 & = & \frac{1}{24 \mu^4} \left( -t^2 + 2 M_{100} t + \Delta - 3 \sqrt{D} \right),
 \nonumber \\
 e_3 & = & \frac{1}{24 \mu^4} \left( 2 t^2 - 4 M_{100} t - 2 \Delta \right),
\eq
where $D$ has been defined in eq.~(\ref{def_D}).
In Euclidean region close to $t=0$ and in the vicinity of the equal mass points the roots are real and ordered as
\bq
 e_2 \le e_3 < 0 < e_1.
\eq
The periods are then
\bq
\label{def_periods}
 \psi_1 =  
 2 \int\limits_{e_2}^{e_3} \frac{dx}{y}
 =
 \frac{4 \mu^2}{D^{\frac{1}{4}}} K\left(k\right),
 & &
 \psi_2 =  
 2 \int\limits_{e_1}^{e_3} \frac{dx}{y}
 =
 \frac{4 i \mu^2}{D^{\frac{1}{4}}} K\left(k'\right),
\eq
where the modulus $k$ and complementary modulus $k'$ are given by
\bq
\label{def_modulus}
 k = \sqrt{\frac{e_3-e_2}{e_1-e_2}},
 & &
 k' = \sqrt{1-k^2} = \sqrt{\frac{e_1-e_3}{e_1-e_2}}.
\eq
$K(x)$ denotes the complete elliptic integral of the first kind
\bq
 K(x)
 & = &
 \int\limits_0^1 \frac{dt}{\sqrt{\left(1-t^2\right)\left(1-x^2t^2\right)}}.
\eq
We will use the standard definition for the ratio $\tau$ of the two periods and the nome $q$. These are given by
\bq
\label{standard_def}
 \tau = i \frac{K\left(k'\right)}{K\left(k\right)},
 & &
 q = e^{i\pi \tau}.
\eq
It is easily verified that $\psi_1$ and $\psi_2$ are solutions of the homogeneous differential equation by inserting $\psi_1$ and $\psi_2$
into the left-hand side of eq.~(\ref{second_order_dgl}).
The Wronski determinant is given by
\bq
\label{def_Wronski}
 W & = &
 \psi_1 \frac{d}{dt} \psi_2 - \psi_2 \frac{d}{dt} \psi_1
 =
 -
 4 \pi i \mu^4 
 \;
 \frac{\left( 3 t^2 - 2 M_{100} t + \Delta \right)}{t\left( t - \mu_1^2 \right)\left( t - \mu_2^2 \right)\left( t - \mu_3^2 \right)\left( t - \mu_4^2 \right)},
\eq
and therefore $\psi_1$ and $\psi_2$ are two independent solutions.
In the limit $t=0$ we have $k=0$ and $k'=1$. The homogeneous solution $\psi_1$ is regular at $t=0$
\bq
 \psi_1\left(t=0\right) & = & \frac{2 \pi \mu^2}{\sqrt{\Delta}},
\eq
while $\psi_2$ has a logarithmic singularity at $t=0$.

%------------------------------------------------------------------------------
\section{Replacing the momentum squared with the nome}
\label{sec:change_variables}

For the full solution of the inhomogeneous differential equation~(\ref{second_order_dgl}) 
we need in addition to the homogeneous solutions $\psi_1$ and $\psi_2$ one special inhomogeneous solution.
Using the variation of the constants, the special inhomogeneous solution
is given by
\bq
\label{inhomogeneous_orig}
 S_{\mathrm{special}} & = &
 \mu^2 \int\limits_{0}^{t} dt_1 \frac{p_3(t_1)}{p_2(t_1) W(t_1)} \left[ - \psi_1(t) \psi_2(t_1) + \psi_2(t) \psi_1(t_1) \right].
\eq
Our aim is here to express this special solution as a combination of the periods $\psi_1$, $\psi_2$ 
and suitable generalisations of polylogarithms.
To this aim we follow \cite{Bloch:2013tra} and we consider the change of variables from $t=p^2$ to the nome $q$, 
defined in eq.~(\ref{standard_def}).
We need to express the variable $t$ as a function of $q$. The required relation can be derived as follows:
From eq.~(\ref{def_modulus}) we find
\bq
 k^2 k'{}^2 & = & - \frac{16 m_1^2 m_2^2 m_3^2 t}{D}.
\eq
With the standard relations 
\bq
 k = 4 \frac{\eta\left(\frac{\tau}{2}\right)^4 \eta\left(2\tau\right)^8}{\eta\left(\tau\right)^{12}},
 \;\;\;\;\;\;
 k' = \frac{\eta\left(\frac{\tau}{2}\right)^8 \eta\left(2\tau\right)^4}{\eta\left(\tau\right)^{12}},
\eq
where $\eta(\tau)$ denotes Dedekind's $\eta$-function
\bq
 \eta\left(\tau\right)
 & = &
 e^{\frac{\pi i \tau}{12}} \prod\limits_{n=1}^\infty \left( 1- e^{2 \pi i n \tau} \right)
 =
 q^{\frac{1}{12}} \prod\limits_{n=1}^\infty \left( 1 - q^{2n} \right),
\eq
one arrives at
\bq
\label{basic_relation_power_series}
 \frac{t}{\left( \mu_1^2 - t \right) \left( \mu_2^2 - t \right) \left( \mu_3^2 - t\right) \left( \mu_4^2 -t \right)}
 & = &
 - \frac{1}{m_1^2 m_2^2 m_3^2}
 \frac{\eta\left(\frac{\tau}{2}\right)^{24}\eta\left(2\tau\right)^{24}}{\eta\left(\tau\right)^{48}}.
\eq
The left-hand side of eq.~(\ref{basic_relation_power_series}) can be viewed as a power series in $t$, the right-hand side
of eq.~(\ref{basic_relation_power_series}) can be viewed as a power series in $q$.
Eq.~(\ref{basic_relation_power_series}) can be used to express $t$ as a power series in $q$.
The first few terms read
\bq
 t & = &
 - q \frac{\Delta^2}{m_1^2 m_2^2 m_3^2}
 - q^2 \frac{\Delta^2}{m_1^4 m_2^4 m_3^4} \left( 4 M_{300} - 4 M_{210} + 16 M_{111} \right)
 \nonumber \\
 & &
 - q^3 \frac{\Delta^2}{m_1^6 m_2^6 m_3^6} \left( 22 M_{600} - 52 M_{510} + 10 M_{420} + 156 M_{411} - 104 M_{321} + 40 M_{330} + 232 M_{222} \right)
 \nonumber \\
 & &
 + {\cal O}\left(q^4\right).
\eq
For the Jacobian of the transformation from $t$ to $q$ we find
\bq
\label{Jacobian_dq_dt}
 dt & = & \frac{1}{i\pi} \frac{\psi_1^2}{W} \frac{dq}{q}.
\eq
With
\bq
\label{subst_psi2}
 \psi_2 & = & \frac{\psi_1}{i\pi} \ln q
\eq
and partial integration
we can express the special inhomogeneous solution of eq.~(\ref{inhomogeneous_orig}) as 
\bq
\label{inhomogeneous}
 S_{\mathrm{special}} 
 & = &
 - \mu^2 \frac{\psi_1\left(q\right)}{\pi^2}
 \int\limits_0^q \frac{dq_1}{q_1}
 \int\limits_0^{q_1} \frac{dq_2}{q_2}
 \;
 \frac{p_3\left(q_2\right) \psi_1\left(q_2\right)^3}{p_2\left(q_2\right) W\left(q_2\right)^2}.
\eq
The full solution is then given by
\bq
\label{full_solution}
 S & = & S_0 + S_{\mathrm{special}}.
\eq

%------------------------------------------------------------------------------
\section{The equal mass case}
\label{sec:equal}

In this section we review the equal mass case $m_1=m_2=m_3=m$, first discussed in \cite{Bloch:2013tra}. 
In this case several equations simplify considerably.
We have a closed expression for the variable $t$ as a function of the nome $q$:
\bq
 t & = & 
 - 9 m^2 
 \frac{\eta\left(\tau\right)^4 \eta\left(\frac{3\tau}{2}\right)^4 \eta\left(6\tau\right)^4}
      {\eta\left(\frac{\tau}{2}\right)^4 \eta\left(2\tau\right)^4 \eta\left(3\tau\right)^4}.
\eq
For the integrand of eq.~(\ref{inhomogeneous}) we find in the equal mass case
\bq
 - \frac{\mu^2}{\pi} \frac{p_3\left(q\right) \psi_1\left(q\right)^3}{p_2\left(q\right) W\left(q\right)^2}
 = 
 3 \sqrt{3} 
 \frac{ \eta\left(\tau\right)^{11} \eta\left(3\tau\right)^{7} }
      { \eta\left(\frac{\tau}{2}\right)^{5} \eta\left(\frac{3\tau}{2}\right) \eta\left(2\tau\right)^{5} \eta\left(6\tau\right) }
 =
 3 \sqrt{3} 
 \sum\limits_{k=1}^\infty k^2 \frac{q^k}{1+\left(-q\right)^k+q^{2k}}.
\eq
Expanding the last term in $q$ and integrating term by term we arrive at
the special solution of the inhomogeneous differential equation
\bq
 S_{\mathrm{special}} 
 & = &
 \frac{3 \psi_1\left(q\right)}{i \pi}
 \sum\limits_{k=1}^\infty 
 \left(-1\right)^k
 \left[ \mathrm{Li}_2\left( r_3 \left(-q\right)^k \right) - \mathrm{Li}_2\left( r_3^{-1} \left(-q\right)^k \right)\right].
\eq
Putting everything together we find that the full solution in the equal mass case can be written as
\bq
\label{full_solution_equal_mass}
 S
 = 
 \frac{3 \psi_1\left(q\right)}{i \pi}
 \left\{
 \frac{1}{2} \left[ \mathrm{Li}_2\left( r_3 \right) - \mathrm{Li}_2\left( r_3^{-1} \right)\right]
 +
 \sum\limits_{k=1}^\infty 
 \left(-1\right)^k
 \left[ \mathrm{Li}_2\left( r_3 \left(-q\right)^k \right) - \mathrm{Li}_2\left( r_3^{-1} \left(-q\right)^k \right)\right]
 \right\}.
\eq

%------------------------------------------------------------------------------
\section{Generalisations of the polylogarithms}
\label{sec:polylogs}

The classical polylogarithms are defined by
\bq
 \mathrm{Li}_n\left(x\right) & = & \sum\limits_{j=1}^\infty \; \frac{x^j}{j^n}.
\eq
Here we consider the following generalisation depending on three variables $x$, $y$, $q$ and two (integer) indices $n$, $m$:
\bq
 \mathrm{ELi}_{n;m}\left(x;y;q\right) & = & 
 \sum\limits_{j=1}^\infty \sum\limits_{k=1}^\infty \; \frac{x^j}{j^n} \frac{y^k}{k^m} q^{j k}.
\eq
This definition is symmetric under the exchange of the pair $(x,n)$ with $(y,m)$.
The two summations are coupled through the variable $q$.
We define the weight of $\mathrm{ELi}_{n;m}(x;y;q)$ to be $w=n+m$.
Note that
\bq
\lefteqn{
 \mathrm{ELi}_{n;m}\left(x;y;-q\right)
 = } & & 
 \nonumber \\
 & &
 \frac{1}{2} \left[
 \mathrm{ELi}_{n;m}\left(x;y;q\right)
 + \mathrm{ELi}_{n;m}\left(x;-y;q\right)
 + \mathrm{ELi}_{n;m}\left(-x;y;q\right)
 - \mathrm{ELi}_{n;m}\left(-x;-y;q\right)
 \right],
\eq
which relates the $\mathrm{ELi}_{n;m}$-function with argument $(-q)$ to a linear combination of $\mathrm{ELi}_{n;m}$-functions
with argument $q$. 
The solution in the equal mass case in eq.~(\ref{full_solution_equal_mass}) can then be written as
\bq
\label{full_solution_equal_mass_ELi}
 S
 =
 \frac{3 \psi_1\left(q\right)}{i \pi}
 \left\{
 \frac{1}{2} \mathrm{Li}_2\left( r_3 \right) - \frac{1}{2} \mathrm{Li}_2\left( r_3^{-1} \right)
 + \mathrm{ELi}_{2;0}\left(r_3;-1;-q\right)
 - \mathrm{ELi}_{2;0}\left(r_3^{-1};-1;-q\right)
 \right\}.
\eq
The combination inside the curly bracket occurs frequently and we introduce for $y\in\{-1,1\}$
the notation
\bq
 \mathrm{E}_{2;0}\left(x;y;q\right)
 & = &
 \frac{1}{i}
 \left[
 \frac{1}{2} \mathrm{Li}_2\left( x \right) 
 - \frac{1}{2} \mathrm{Li}_2\left( x^{-1} \right)
 + \mathrm{ELi}_{2;0}\left(x;y;q\right)
 - \mathrm{ELi}_{2;0}\left(x^{-1};y^{-1};q\right)
 \right].
\eq
We call the function $\mathrm{E}_{2;0}\left(x;y;q\right)$ an elliptic dilogarithm.
We remark that different definitions of elliptic polylogarithms have been considered in \cite{Beilinson:1994,Levin:1997,Levin:2007,Brown:2011,Wildeshaus}.
With the notation above we may write the solution in the equal mass case as
\bq
\label{full_solution_equal_mass_E}
 S
 & = &
 3 
 \; 
 \frac{\psi_1\left(q\right)}{\pi}
 \;
 \mathrm{E}_{2;0}\left(r_3;-1;-q\right).
\eq

%------------------------------------------------------------------------------
\section{The unequal mass case}
\label{sec:unequal_mass_case}

The final result for the two-loop sunrise integral with arbitrary masses in two space-time dimensions 
is almost as simple as the result in the equal mass case given in eq.~(\ref{full_solution_equal_mass_E}).
The result in the unequal mass case can be written as
\bq
\label{full_solution_unequal_mass_E}
 S
 & = &
 \frac{\psi_1\left(q\right)}{\pi}
 \left[
 \mathrm{E}_{2;0}\left(w_1\left(q\right);-1;-q\right)
 +
 \mathrm{E}_{2;0}\left(w_2\left(q\right);-1;-q\right)
 +
 \mathrm{E}_{2;0}\left(w_3\left(q\right);-1;-q\right)
 \right].
\eq
Eq.~(\ref{full_solution_unequal_mass_E}) differs from eq.~(\ref{full_solution_equal_mass_E}) only in the arguments
$w_1$, $w_2$, $w_3$ of the elliptic dilogarithms.
These arguments are given by
\bq
\label{def_arguments_w_i}
 w_i = e^{i \beta_i},
 \;\;\;\;
 \beta_i = \pi \frac{F\left(u_i,k\right)}{K\left(k\right)},
 \;\;\;\;
 u_i = \sqrt{\frac{e_1-e_2}{x_{j,k}-e_2}},
 \;\;\;\;
 x_{j,k} = e_3 + \frac{m_j^2 m_k^2}{\mu^4}.
\eq
In the definition of $u_i$ we used the convention that $(i,j,k)$ is a permutation of $(1,2,3)$.
In the definition of $\beta_i$ the incomplete elliptic integral of the first kind appears, defined by
\bq
 F\left(z,x\right)
 & = &
 \int\limits_0^z \frac{dt}{\sqrt{\left(1-t^2\right)\left(1-x^2t^2\right)}}.
\eq
The quantities $w_i$ and $\beta_i$ depend in general on the masses $m_i$ and the nome $q$.
In the equal mass case $m_1=m_2=m_3=m$ they are independent of $q$ and we have
\bq
 w_i\left(q\right) = r_3, & & \beta_i\left(q\right) = \frac{2 \pi}{3}.
\eq
At zero momentum squared ($t=0$, corresponding to $q=0$, the masses not necessarily equal) we have
\bq
 w_i\left(0\right) = v_i, & & \beta_i\left(0\right) = \alpha_i.
\eq
As a consequence
\bq
 \mathrm{E}_{2;0}\left(w_i\left(0\right);-1;0\right)
 & = &
 \frac{1}{2i}
 \left[ \mathrm{Li}_2\left( v_i \right) - \mathrm{Li}_2\left( v_i^{-1} \right)
 \right],
\eq
and eq.~(\ref{full_solution_unequal_mass_E}) reduces for $q=0$ (which corresponds to $t=0$) to eq.~(\ref{boundary_2}).

The correctness of eq.~(\ref{full_solution_unequal_mass_E}) is easily verified by comparing the $q$-expansion of 
eq.~(\ref{full_solution_unequal_mass_E}) with the $q$-expansion obtained 
from eq.~(\ref{full_solution}) together with eq.~(\ref{inhomogeneous}).

%------------------------------------------------------------------------------
\section{Geometric interpretation}
\label{sec:geometric_interpretation}

The arguments $w_1$, $w_2$, $w_3$ appearing in the elliptic dilogarithms have a simple geometric interpretation.
We recall that an elliptic curve can be represented in different ways.
Starting from the original elliptic curves $E_i$, which were given as the cubic curve ${\mathcal F}=0$ together
with the choice $O=P_i$ as origin, we can go over to the elliptic curve $\hat{E}$ in Weierstrass normal form.
From there we can move on to a torus ${\mathbb C}/\Lambda$, where $\Lambda$ is a lattice defined by the
periods of the elliptic curve $\hat{E}$.
Finally, we can transform from the torus to the Jacobi uniformization ${\mathbb C}^\ast/{q^{2 {\mathbb Z}}}$.
We thus have a sequence of mappings
\bq
\label{chain_of_mappings}
\begin{CD}
 E_i @>>> \hat{E} @>>> {\mathbb C}/\Lambda @>>> {\mathbb C}^\ast/{q^{2 {\mathbb Z}}}.
\end{CD}
\eq
Distinguished points are the points on the intersection of the cubic curve ${\mathcal F}=0$ 
with the domain of integration $\sigma$.
There are three intersection points, which we recall from eq.~(\ref{intersection_F_sigma})
\bq
 P_1 = \left[1:0:0\right], 
 \;\;\;
 P_2 = \left[0:1:0\right], 
 \;\;\;
 P_3 = \left[0:0:1\right].
\eq
As a side-remark we emphasise
that $\sigma$ is defined in eq.~(\ref{def_sigma}) by the condition $x_i \ge 0$, therefore
the points
\bq
 P_4 = \left[ 0 : m_3^2 : - m_2^2 \right],
 \;\;\;
 P_5 = \left[ - m_3^2 : 0 : m_1^2 \right],
 \;\;\;
 P_6 = \left[ m_2^2 : - m_1^2 : 0 \right]
\eq
are not in $\sigma$.
The points $P_1$, $P_2$ and $P_3$ determine the arguments $w_1$, $w_2$, $w_3$ of the elliptic dilogarithms as follows:
The elliptic curve $E_j$ is defined by the cubic curve ${\mathcal F}=0$ and the choice $O=P_j$ as origin.
We denote by $Q_{i,j}$ the image of the point $P_i \in E_j$ on $\hat{E}$.
The explicit formulae for the transformation from $E_j$ to $\hat{E}$ are given in appendix~\ref{trafo_to_WNF}.
By construction we have
\bq
 Q_{i,i} & = & \left[ 0 : 1 : 0 \right].
\eq
Of particular relevance are the six points $Q_{i,j}$ with $i \neq j$.
These are given by
\bq
\label{points_on_E_hat}
 Q_{1,2} & = &
 \left[e_3+\frac{m_1^2 m_2^2}{\mu^4}:\frac{m_1^2 m_2^2\left(t-m_1^2-m_2^2+m_3^2\right)}{\mu^6}:1\right],
 \nonumber \\
 Q_{2,3} & = & 
 \left[e_3+\frac{m_2^2 m_3^2}{\mu^4}:\frac{m_2^2 m_3^2\left(t+m_1^2-m_2^2-m_3^2\right)}{\mu^6}:1\right],
 \nonumber \\
 Q_{3,1} & = &
 \left[e_3+\frac{m_1^2 m_3^2}{\mu^4}:\frac{m_1^2 m_3^2\left(t-m_1^2+m_2^2-m_3^2\right)}{\mu^6}:1\right],
 \nonumber \\
 Q_{2,1} & = &
 \left[e_3+\frac{m_1^2 m_2^2}{\mu^4}:-\frac{m_1^2 m_2^2\left(t-m_1^2-m_2^2+m_3^2\right)}{\mu^6}:1\right],
 \nonumber \\
 Q_{3,2} & = &
 \left[e_3+\frac{m_2^2 m_3^2}{\mu^4}:-\frac{m_2^2 m_3^2\left(t+m_1^2-m_2^2-m_3^2\right)}{\mu^6}:1\right],
 \nonumber \\
 Q_{1,3} & = & 
 \left[e_3+\frac{m_1^2 m_3^2}{\mu^4}:-\frac{m_1^2 m_3^2\left(t-m_1^2+m_2^2-m_3^2\right)}{\mu^6}:1\right].
\eq
We denote the coordinates of $Q_{i,j}$ by $[x_{i,j} : y_{i,j} : 1 ]$. From eq.~(\ref{points_on_E_hat}) it follows that
\bq
 x_{i,j} = x_{j,i},
 & &
 y_{i,j} = - y_{j,i},
\eq
and therefore we may write
\bq
 Q_{i,j} & = & - Q_{j,i}
\eq
with respect to the addition on $\hat{E}$.

As a side-remark we mention that in the equal mass case $m_1=m_2=m_3=m$ the points $Q_{i,j}$ are given by
\bq
 Q_{i,j} & = & 
 \left[
 \frac{t^2-6m^2 t+ 9 m^4}{12 \mu^4} 
 :
 \pm \frac{m^4 \left(t-m^2\right)}{\mu^6}
 :
 1
 \right].
\eq
In the equal mass case these points are torsion points of order 3.
This is no longer true in the unequal mass case.
However, the torsion property is not essential for our analysis.

We continue with the general unequal mass case. 
The periods $\psi_1$ and $\psi_2$ (defined in eq.~(\ref{def_periods})) of the elliptic curve $\hat{E}$ define
a lattice.
It will be convenient to work with a lattice, where the first generator is normalised to $1$.
We take $\Lambda$ to be the lattice generated by $1$ and $\psi_2/\psi_1$.
The set ${\mathbb C}/\Lambda$ is a torus, and we will denote the complex coordinate on the torus by $\hat{z}$.
There is a mapping from $\hat{E}$ to ${\mathbb C}/\Lambda$, given by
\bq
 \left[x:y:1\right] 
 & \rightarrow &
 \hat{z} =
 \frac{1}{\psi_1} 
 \int\limits_{x}^\infty 
 \frac{d\tilde{x}}{\sqrt{4\left(\tilde{x}-e_1\right)\left(\tilde{x}-e_2\right)\left(\tilde{x}-e_3\right)}}.
\eq
The integral is an incomplete elliptic integral of the first kind. Transforming this integral into the standard form, we find 
that the points $Q_{1,2}$, $Q_{2,3}$ and $Q_{3,1}$ are mapped to
\bq
 Q_{j,k}
 & \rightarrow &
 \hat{z}_i =
 \frac{1}{2} \frac{F\left(u_i,k\right)}{K\left(k\right)},
 \;\;\;
 u_i = \sqrt{\frac{e_1-e_2}{x_{j,k}-e_2}}.
\eq
Here we used the convention that the triple $(i,j,k)$ is a cyclic permutation of $(1,2,3)$.
The points $Q_{2,1}$, $Q_{3,2}$ and $Q_{1,3}$ are mapped to
\bq
 Q_{j,k}
 & \rightarrow &
 - \hat{z}_i.
\eq
From the torus ${\mathbb C}/\Lambda$ we have a mapping to the 
Jacobi uniformization ${\mathbb C}^\ast/{q^{2 {\mathbb Z}}}$ given by
\bq
 \hat{z} & \rightarrow & 
 w = e^{2 \pi i \hat{z}}.
\eq
The points $\hat{z}_1$, $\hat{z}_2$ and $\hat{z}_3$ are transformed to
\bq
 \hat{z}_i & \rightarrow & w_i,
 \;\;\;\;\;\; i \in \{1,2,3\},
\eq
with $w_i$ given in eq.~(\ref{def_arguments_w_i}),
while the points
$(-\hat{z}_1)$, $(-\hat{z}_2)$ and $(-\hat{z}_3)$ are transformed to
\bq
 -\hat{z}_i & \rightarrow & w_i^{-1}.
\eq
Therefore we see that the arguments $w_1$, $w_2$ and $w_3$ (as well as $w_1^{-1}$, $w_2^{-1}$ and $w_3^{-1}$)
of the elliptic dilogarithms
are the images of the intersection points of ${\mathcal F}=0$ with the integration region $\sigma$ 
which are not chosen as the origin of the elliptic curve
under the chain of mappings given in eq.~(\ref{chain_of_mappings}).

% ------------------------------------------------------------------------------------------
\section{Conclusions}
\label{sec:conclusions}

In this paper we presented the result for the 
two-loop sunrise integral with arbitrary non-zero masses in two space-time dimensions
in terms of elliptic dilogarithms.
The result is given in eq.~(\ref{full_solution_unequal_mass_E}).
Remarkably, the result is almost as simple as the corresponding result in the equal mass case,
only the arguments of the elliptic dilogarithms are modified.
We would like to emphasise the geometric interpretation of the ingredients in our final result.
The sunrise integral is given as the product of a period of an elliptic curve with a sum of elliptic dilogarithms.
The arguments of the elliptic dilogarithms are 
the images of the intersection points of the variety defined by the graph polynomial
with the integration region under a chain of mappings defined by eq.~(\ref{chain_of_mappings}).

\subsection*{Acknowledgements}

C.B. thanks Humboldt University for support and hospitality.

% ------------------------------------------------------------------------------------------
\begin{appendix}

\section{The Weierstrass normal form}
\label{trafo_to_WNF}

In this appendix we give the explicit formulae to transform the elliptic curve
\bq
 E_i & : & {\cal F} = 0 \;\;\;\mbox{with} \;\;\; O = P_i,
\eq
into the Weierstrass normal form
\bq
 \hat{E} & : &  y^2 z = 4 x^3 - g_2 x z^2 - g_3 z^3.
\eq
The three elliptic curves $E_1$, $E_2$, $E_3$ differ by the choice of the origin.
For the curve $E_i$ we choose the point $O=P_i$ as origin, where
\bq
 P_1 = \left[1:0:0\right], 
 \;\;\;
 P_2 = \left[0:1:0\right], 
 \;\;\;
 P_3 = \left[0:0:1\right].
\eq
We first consider the elliptic curve $E_3$ corresponding to the choice
\bq
 O & = & P_3.
\eq
We set
\bq
 x & = &
 \frac{1}{\mu^{10}}
 \left( m_1^2 x_1 + m_2^2 x_2 \right)
 \left[
         m_1^4 \left( t^2 - 2 M_{100} t - \Delta + 12 m_1^2 m_3^2 \right) x_1^2
 \right. \nonumber \\
 & & \left.
       + m_2^4 \left( t^2 - 2 M_{100} t - \Delta + 12 m_2^2 m_3^2 \right) x_2^2
 \right. \nonumber \\
 & & \left.
       + 2 m_1^2 m_2^2 \left( t^2 - 2 M_{100} t - \Delta + 6 m_1^2 m_3^2 + 6 m_2^2 m_3^2 \right) x_1 x_2
 \right. \nonumber \\
 & & \left.
       + 12 m_1^2 m_3^2 \left(-m_1^2 t+m_1^4+m_2^4-2 m_1^2 m_2^2+m_1^2 m_3^2 \right) x_1 x_3
 \right. \nonumber \\
 & & \left.
       + 12 m_2^2 m_3^2 \left(-m_2^2 t+m_1^4+m_2^4-2 m_1^2 m_2^2+m_2^2 m_3^2 \right) x_2 x_3
       + 12 m_3^4 \left(m_1^2-m_2^2 \right)^2 x_3^2
 \right],
 \nonumber \\
 y & = &
 \frac{12 m_3^2}{\mu^{12}}
 \left[
          m_1^8 \left(-t+m_1^2-m_2^2+m_3^2\right) x_1^3
          +m_2^8 \left(t+m_1^2-m_2^2-m_3^2\right) x_2^3
 \right. \nonumber \\
 & & \left.
          +m_1^4 m_2^2 \left(-2 m_1^2 t+m_2^2 t+2 m_1^4-m_2^4-m_1^2 m_2^2-m_2^2 m_3^2+2 m_3^2 m_1^2\right) x_1^2 x_2
 \right. \nonumber \\
 & & \left.
          +m_1^2 m_2^4 \left(-m_1^2 t+2 m_2^2 t+m_1^4-2 m_2^4+m_1^2 m_2^2-2 m_2^2 m_3^2+m_3^2 m_1^2\right) x_1 x_2^2
 \right. \nonumber \\
 & & \left.
          +m_1^4 \left(
                       m_1^2 t^2
                       - 2 m_1^4 t + m_1^2 m_2^2 t - 2 m_1^2 m_3^2 t + m_2^4 t    
                       + m_1^6 - m_2^6
                       + 4 m_1^4 m_3^2 - 3 m_1^4 m_2^2 
                       + 3 m_1^2 m_2^4 
 \right. \right. \nonumber \\
 & & \left. \left.
                       + m_1^2 m_3^4 - 3 m_1^2 m_2^2 m_3^2  
                       - m_2^4 m_3^2 
                 \right) x_1^2 x_3
 \right. \nonumber \\
 & & \left.
          +m_2^4 \left(
                       - m_2^2 t^2
                       - m_1^4 t
                       - m_1^2 m_2^2 t
                       + 2 m_2^4 t
                       + 2 m_2^2 m_3^2 t 
                       + m_1^6
                       - m_2^6
                       - 3 m_1^4 m_2^2
                       + m_1^4 m_3^2
                       + 3 m_1^2 m_2^4
 \right. \right. \nonumber \\
 & & \left. \left.
                       + 3 m_1^2 m_2^2 m_3^2
                       - m_2^2 m_3^4
                       - 4 m_2^4 m_3^2
                 \right) x_2^2 x_3
 \right. \nonumber \\
 & & \left.
          +m_1^2 m_2^2 \left(m_1^2-m_2^2\right) 
           \left(
                 + t^2
                 - 3 m_1^2 t
                 - 3 m_2^2 t
                 - 2 m_3^2 t
                 + 2 m_1^4
                 + 2 m_2^4
                 + m_3^4
                 - 4 m_1^2 m_2^2
                 + 5 m_1^2 m_3^2 
 \right. \right. \nonumber \\
 & & \left. \left.
                 + 5 m_2^2 m_3^2
            \right) x_1 x_2 x_3
 \right. \nonumber \\
 & & \left.
          +m_1^2 m_3^2 \left(m_1^2-m_2^2\right) 
           \left(
                 - 3 m_1^2 t
                 - m_2^2 t
                 + 3 m_1^4
                 + 3 m_2^4
                 - 6 m_1^2 m_2^2
                 + 3 m_1^2 m_3^2 
                 + m_2^2 m_3^2
           \right) x_1 x_3^2
 \right. \nonumber \\
 & & \left.
          +m_2^2 m_3^2 \left(m_1^2-m_2^2\right) 
           \left(
                 - m_1^2 t
                 - 3 m_2^2 t
                 + 3 m_1^4
                 + 3 m_2^4
                 - 6 m_1^2 m_2^2
                 + m_1^2 m_3^2 
                 + 3 m_2^2 m_3^2
           \right) x_2 x_3^2
 \right. \nonumber \\
 & & \left.
          +2 m_3^4 \left(m_1^2-m_2^2\right)^3 x_3^3
 \right],
 \nonumber \\
 z & = &
 \frac{12}{\mu^6} \left(m_1^2 x_1+m_2^2 x_2 \right)^3.
\eq
The inverse transformation is given by
\bq
 x_1 & = &
 \frac{m_3^2}{2 \mu^2} z
 \left[ x - \frac{1}{12 \mu^4} \left(t^2-2 M_{100} t-\Delta+12 m_2^2 m_3^2\right) z\right]
 \nonumber \\
 & &
 \left[
           y
           -\frac{1}{\mu^2} \left(t-m_1^2+m_2^2-m_3^2\right) x
           +\frac{1}{12 \mu^6} \left(t-m_1^2+m_2^2-m_3^2\right) \left(t^2-2 M_{100} t-\Delta\right) z
 \right],
 \nonumber \\
 x_2 & = &
 \frac{m_3^2}{2 \mu^2} z
 \left[ x - \frac{1}{12 \mu^4} \left(t^2-2 M_{100} t-\Delta+12 m_1^2 m_3^2\right) z\right]
 \nonumber \\
 & &
 \left[
           -y
           -\frac{1}{\mu^2}\left(t+m_1^2-m_2^2-m_3^2\right) x
           +\frac{1}{12 \mu^6} \left(t+m_1^2-m_2^2-m_3^2\right) \left(t^2-2 M_{100} t-\Delta\right) z
 \right],
 \nonumber \\
 x_3 & = &
 \left[x-\frac{1}{12 \mu^4} \left(t^2-2 M_{100} t-\Delta\right) z\right]
 \left[x-\frac{1}{12 \mu^4} \left(t^2-2 M_{100} t-\Delta+12 m_1^2 m_3^2\right) z\right]
 \nonumber \\
 & &
 \left[x-\frac{1}{12 \mu^4} \left(t^2-2 M_{100} t-\Delta+12 m_2^2 m_3^2\right) z\right].
\eq
The above change of variables transforms the points $P_1$, $P_2$ and $P_3$ to
\bq
 P_1
 & \rightarrow &
 Q_{1,3}=\left[e_3+\frac{m_1^2 m_3^2}{\mu^4}:-\frac{m_1^2 m_3^2\left(t-m_1^2+m_2^2-m_3^2\right)}{\mu^6}:1\right],
 \nonumber \\
 P_2
 & \rightarrow &
 Q_{2,3}=\left[e_3+\frac{m_2^2 m_3^2}{\mu^4}:\frac{m_2^2 m_3^2\left(t+m_1^2-m_2^2-m_3^2\right)}{\mu^6}:1\right],
 \nonumber \\
 P_3
 & \rightarrow &
 Q_{3,3}=\left[0:1:0\right],
\eq
where $e_3$ has been defined in eq.~(\ref{def_roots}).
We recall that we first considered the elliptic curve $E_3$ corresponding to the choice $O=P_3$ as origin,
and consequently $P_3$ is transformed to $Q_{3,3}=[0:1:0]$, the point of origin of the Weierstrass normal form.
Corresponding formulae for the change of variables starting from the elliptic curves $E_1$ or $E_2$ 
are obtained by a simultaneous (cyclic) permutation of $(x_1,x_2,x_3)$ and
$(m_1,m_2,m_3)$.

\section{Changing the variable from the momentum squared to the nome}

In this appendix we give a few details on the change of the variable from the momentum squared $t$ to the nome $q$.
We start with the Jacobian 
\bq
\label{Jacobian_dq_dt_v2}
 \frac{dq}{dt} & = &  i \pi
 \frac{q}{\psi_1^2}
 W.
\eq
in eq.~(\ref{Jacobian_dq_dt}). This equation is derived as follows:
From the definition $q=\exp(i\pi\tau)$, $\tau=iK(k')/K(k)$  we have
\bq
 \frac{dq}{dt} & = & i \pi q \frac{d\tau}{dk} \frac{dk}{dt}.
\eq
Using the relation $k^2+k'{}^2=1$ and the Legendre relation 
\bq
 K\left(k\right) E\left(k'\right) + E\left(k\right) K\left(k'\right) - K\left(k\right) K\left(k'\right) & = & \frac{\pi}{2}
\eq
we find for $d\tau/dk$:
\bq
 \frac{d\tau}{dk}
 & = & 
 -\frac{i \pi}{2 k k'{}^2 \left(K\left(k\right)\right)^2}
\eq
In order to determine $dk/dt$ we first introduce the quantities
\bq
 Z_1 & = & e_3 - e_2 = \frac{1}{8 \mu^4} \left( t^2 - 2 M_{100} t - 2 M_{110} + M_{200} + \sqrt{D} \right),
 \nonumber \\
 Z_2 & = & e_1 - e_3 = \frac{1}{8 \mu^4} \left( -t^2 + 2 M_{100} t + 2 M_{110} - M_{200} + \sqrt{D} \right),
 \nonumber \\
 Z_3 & = & e_1 - e_2 = \frac{1}{4 \mu^4} \sqrt{D}.
\eq
We have
\bq
 k = \sqrt{\frac{Z_1}{Z_3}},
 &&
 k' = \sqrt{\frac{Z_2}{Z_3}}.
\eq
We then find
\bq
\frac{dk}{dt}
 & = &
 \frac{1}{2k Z_3^2} \left( Z_2 \frac{dZ_1}{dt} - Z_1 \frac{dZ_2}{dt} \right).
\eq
With
\bq
 Z_2 \frac{dZ_1}{dt} - Z_1 \frac{dZ_2}{dt}
 & = &
 - \frac{1}{4 Z_3 \mu^{12}} m_1^2 m_2^2 m_3^2 \left( 3 t^2 - 2 M_{100} t + 2 M_{110} - M_{200} \right),
 \nonumber \\
 Z_1 Z_2 
 & = & 
 - \frac{m_1^2 m_2^2 m_3^2 t}{\mu^8},
\eq
and the definition of the Wronski determinant in eq.~(\ref{def_Wronski}) one arrives at eq.~(\ref{Jacobian_dq_dt_v2}).

Next we consider the change of the integration variable from eq.~(\ref{inhomogeneous_orig}) to eq.~(\ref{inhomogeneous}).
We start from eq.~(\ref{inhomogeneous_orig})
\bq
 S_{\mathrm{special}} & = &
 - \psi_1(t) \int\limits_{0}^{t} dt_1 \frac{\mu^2 p_3(t_1)}{p_2(t_1) W(t_1)} \psi_2(t_1)
 + \psi_2(t) \int\limits_{0}^{t} dt_1 \frac{\mu^2 p_3(t_1)}{p_2(t_1) W(t_1)} \psi_1(t_1)
\eq
Changing the integration variable from $t_1$ to $q_1$, using eq.~(\ref{subst_psi2}) and setting
\bq
 f(q) & = & \frac{\mu^2 p_3(q) \left(\psi_1(q)\right)^3}{\pi^2 p_2(q) \left(W(q)\right)^2}
\eq
we arrive at
\bq
 S_{\mathrm{special}} & = &
 \psi_1(q) \int\limits_0^q \frac{dq_1}{q_1} f\left(q_1\right) \ln\left(q_1\right)
 -
 \psi_1(q) \ln(q) \int\limits_0^q \frac{dq_1}{q_1} f\left(q_1\right).
\eq
We can now use partial integration with
\bq
 F(q)=\int\limits_0^q \frac{dq_1}{q_1} f\left(q_1\right),
 & &
 G(q)=\ln\left(q\right),
\eq
with the result
\bq
 S_{\mathrm{special}} & = &
 - \psi_1(q) \int\limits_0^q \frac{dq_1}{q_1} F\left(q_1\right),
\eq
which agrees with eq.~(\ref{inhomogeneous}).

%------------------------------------------------------------------------------
\section{The expansion in the nome}
\label{sec:nome_expansion}

In this appendix we give information on how to expand 
eq.~(\ref{full_solution}) (in combination with eq.~(\ref{inhomogeneous})) 
and eq.~(\ref{full_solution_unequal_mass_E}) in the variable $q$.
We start with the first case and write eq.~(\ref{full_solution}) as
\bq
 S
 & = &
 \frac{\psi_1\left(q\right)}{\pi} R\left(q,m_1,m_2,m_3\right),
\eq
with
\bq
 R\left(q,m_1,m_2,m_3\right)
 & = &
 R_0\left(m_1,m_2,m_3\right)
 +
 R_1\left(q,m_1,m_2,m_3\right)
\eq
and
\bq
 R_0 & = &
 \frac{1}{2i}
 \left[
        \mathrm{Li}_2\left( v_1 \right) + \mathrm{Li}_2\left( v_2 \right) + \mathrm{Li}_2\left( v_3 \right) 
      - \mathrm{Li}_2\left( v_1^{-1} \right) - \mathrm{Li}_2\left( v_2^{-1} \right) - \mathrm{Li}_2\left( v_3^{-1} \right) 
 \right],
 \nonumber \\
 R_1
 & = &
 - \frac{\mu^2}{\pi} 
 \int\limits_0^q \frac{dq_1}{q_1}
 \int\limits_0^{q_1} \frac{dq_2}{q_2}
 \;
 \frac{p_3\left(q_2\right) \psi_1\left(q_2\right)^3}{p_2\left(q_2\right) W\left(q_2\right)^2},
\eq
For the $q$-expansion of $R_1$ we write
\bq
\label{q_expansion_R1}
 R_1 & = &
 \sum\limits_{n=1}^\infty 
 q^n
 \left[ 
        c_n
      + d_{1,n} \ln\frac{m_1^2}{\mu^2}
      + d_{2,n} \ln\frac{m_2^2}{\mu^2}
      + d_{3,n} \ln\frac{m_3^2}{\mu^2}
 \right].
\eq
The coefficients $c_n$ are symmetric in $m_1$, $m_2$ and $m_3$.
The coefficients $d_{i,n}$ satisfy
\bq
\label{d_relation}
 d_{1,n} + d_{2,n} + d_{3,n} & = & 0.
\eq
Due to eq.~(\ref{d_relation}) we have
\bq
 \sum\limits_{i=1}^3
 d_{i,n} \ln\frac{m_i^2}{\mu^2}
 & = &
 -
 \sum\limits_{i=1}^3
 d_{i,n}
 \left[ \mathrm{Li}_1\left(v_i\right) + \mathrm{Li}_1\left(v_i^{-1}\right) \right]. 
\eq
In order to obtain the $q$-expansion of $R_1$, we first express the variable $t$ as a power series in $q$,
as outlined after eq.~(\ref{basic_relation_power_series}).
The $q$-expansion of $R_1$ is then easily obtained.
The first few coefficients of the $q$-expansion in eq.~(\ref{q_expansion_R1}) are given by
\bq
 c_1
 & = & 
 \frac{1}{i}
 \left[ v_1 + v_2 + v_3 - v_1^{-1} - v_2^{-1} - v_3^{-1} \right],
 \\
 c_2
 & = &
 \frac{1}{i}
 \left[ \frac{7}{4} \left( v_1^2 + v_2^2 + v_3^2 - v_1^{-2} - v_2^{-2} - v_3^{-2} \right)
        + 3 \left( v_1 + v_2 + v_3 - v_1^{-1} - v_2^{-1} - v_3^{-1} \right) \right],
 \nonumber \\
 c_3
 & = &
 \frac{1}{i}
 \left[ 
          \frac{55}{9} \left( v_1^3 + v_2^3 + v_3^3 - v_1^{-3} - v_2^{-3} - v_3^{-3} \right)
        + \frac{32}{3} \left( v_1^2 + v_2^2 + v_3^2 - v_1^{-2} - v_2^{-2} - v_3^{-2} \right)
 \right. \nonumber \\
 & & \left.
        + \frac{13}{3} \left( v_1 + v_2 + v_3 - v_1^{-1} - v_2^{-1} - v_3^{-1} \right) 
        + \frac{11}{3} \left( v_1^3 v_2 + v_2^3 v_3 + v_3^3 v_1 + v_2^3 v_1 + v_1^3 v_3 + v_3^3 v_2
 \right. \right. \nonumber \\
 & & \left. \left.
                            - v_1^{-3} v_2^{-1} - v_2^{-3} v_3^{-1} - v_3^{-3} v_1^{-1} - v_2^{-3} v_1^{-1} - v_1^{-3} v_3^{-1} - v_3^{-3} v_2^{-1} \right)
 \right],
 \nonumber
\eq
and
\bq
 d_{1,1}
 & = & 
 \frac{1}{i}
 \left[ - 2 v_1 + v_2 + v_3 + 2 v_1^{-1} - v_2^{-1} - v_3^{-1} \right],
 \\
 d_{1,2}
 & = &
 \frac{1}{i}
 \left[ \frac{7}{2} \left( -2 v_1^2 + v_2^2 + v_3^2 + 2 v_1^{-2} - v_2^{-2} - v_3^{-2} \right)
        + 3 \left( -2 v_1 + v_2 + v_3 + 2 v_1^{-1} - v_2^{-1} - v_3^{-1} \right) 
 \right. \nonumber \\
 & & \left.
        + \frac{7}{2} \left( 
                             - v_1^2 v_2 - v_1^2 v_3 + v_2^2 v_3 + v_3^2 v_2   
                             + v_1^{-2} v_2^{-1} + v_1^{-2} v_3^{-1} - v_2^{-2} v_3^{-1} - v_3^{-2} v_2^{-1}  
                      \right)
\right],
 \nonumber \\
 d_{1,3}
 & = &
 \frac{1}{i}
 \left[ 
          \frac{55}{3} \left( -2 v_1^3 + v_2^3 + v_3^3 + 2 v_1^{-3} - v_2^{-3} - v_3^{-3} \right)
        + 36 \left( -2 v_1^2 + v_2^2 + v_3^2 + 2 v_1^{-2} - v_2^{-2} 
 \right. \right. \nonumber \\
 & & \left. \left.
                    - v_3^{-2} \right)
        - 3 \left( -2 v_1 + v_2 + v_3 + 2 v_1^{-1} - v_2^{-1} - v_3^{-1} \right) 
        + 25 \left( 
                             - v_1^2 v_2 - v_1^2 v_3 + v_2^2 v_3 + v_3^2 v_2   
 \right. \right. \nonumber \\
 & & \left. \left.
                             + v_1^{-2} v_2^{-1} + v_1^{-2} v_3^{-1} - v_2^{-2} v_3^{-1} - v_3^{-2} v_2^{-1}  
                      \right)
        - 55 \left( v_1^3 v_2 + v_1^3 v_3 - v_1^{-3} v_2^{-1} - v_1^{-3} v_3^{-1} \right)
 \right. \nonumber \\
 & & \left.
        + 44 \left( v_2^3 v_3 + v_3^3 v_2 - v_2^{-3} v_3^{-1} - v_3^{-3} v_2^{-1} \right)
        + 11 \left( v_2^3 v_1 + v_3^3 v_1 - v_2^{-3} v_1^{-1} - v_3^{-3} v_1^{-1} \right)
 \right].
 \nonumber
\eq
The coefficients $d_{2,1}$, $d_{2,2}$, $d_{2,3}$ as well as $d_{3,1}$, $d_{3,2}$, $d_{3,3}$ are obtained
by a cyclic permutation of $(v_1,v_2,v_3)$.

In order to obtain the $q$-expansion of eq.~(\ref{full_solution_unequal_mass_E})
we need the $q$-expansion of all quantities appearing in eq.~(\ref{def_arguments_w_i}).
The $q$-expansion of the variables $u_i$ is easily obtained. Writing
\bq
 u_i & = & \sum\limits_{n=0}^\infty u_{i,n} q^n
\eq
one finds for the first few terms
\bq
 u_{1,0} = \frac{\sqrt{\Delta}}{2 m_2 m_3},
 & &
 \sqrt{1-u_{1,0}^2} = \frac{\delta_1}{2 m_2 m_3},
\eq
and
\bq
 u_{1,1} & = & 
 u_{1,0}
 \frac{\delta_1}{m_1^2 m_2^2 m_3^2} \left( - 3 m_1^4 + m_2^4 + m_3^4 + 2 m_1^2 m_2^2 + 2 m_1^2 m_3^2 - 2 m_2^2 m_3^2 \right),
 \\
 u_{1,2} & = &
 u_{1,0}
 \frac{2}{m_1^4 m_2^4 m_3^4}
 \left[
 8 m_1^{12}
 - 26 m_1^{10} \left( m_2^2 + m_3^2 \right)
 + m_1^8 \left( 26 m_2^4 + 26 m_3^4 + 55 m_2^2 m_3^2 \right)
 \right. \nonumber \\
 & & \left.
 - 4 m_1^6 \left( m_2^2 + m_3^2 \right) \left( m_2^4 + m_3^4 + 8 m_2^2 m_3^2 \right)
 + m_1^4 \left( -4 m_2^8 -4 m_3^8 + 6 m_2^6 m_3^2 + 6 m_2^2 m_3^6 
 \right. \right. \nonumber \\
 & & \left. \left.
 + 12 m_2^4 m_3^4 \right)
 - 2 m_1^2 \left( m_2^2 + m_3^2 \right) \left( m_2^2 - m_3^2 \right)^4
 + \left( m_2^2 - m_3^2 \right)^4 \left( 2 m_2^4 + 2 m_3^4 + 3 m_2^2 m_3^2 \right) 
 \right].
 \nonumber
\eq
It is a little bit more tricky to obtain the $q$-expansion of $\beta_i$.
We write
\bq
\label{q_expansion_beta_i}
 \beta_i & = & \sum\limits_{n=0}^\infty \beta_{i,n} q^n.
\eq
The $q$-expansion for $\beta_i$ is obtained as follows:
We note that
\bq
 a_i & = & F\left(u_i,k\right)
\eq
is equivalent to
\bq
 u_i & = & \mathrm{sn}\left(a_i,k\right).
\eq
The Jacobi elliptic function $\mathrm{sn}(a_i,k)$ can be expressed in terms of Jacobi $\theta$-functions:
\bq
 \mathrm{sn}\left(a_i,k\right)
 & = &
 \frac{\theta_3\left(0,q\right)\theta_1\left(b_i,q\right)}{\theta_2\left(0,q\right)\theta_4\left(b_i,q\right)},
 \;\;\;
 b_i = \frac{a_i}{\theta_3^2\left(0,q\right)}.
\eq
We further note that
\bq
 K(k) & = & \frac{\pi}{2} \theta_3^2\left(0,q\right)
\eq
and hence
\bq
 \beta_i & = & 2 b_i.
\eq
We therefore have
\bq
\label{helper_eq_beta_i}
 u_i
 & = &
 \frac{\theta_3\left(0,q\right)\theta_1\left(\frac{\beta_i}{2},q\right)}{\theta_2\left(0,q\right)\theta_4\left(\frac{\beta_i}{2},q\right)}.
\eq
We can now insert eq.~(\ref{q_expansion_beta_i}) into eq.~(\ref{helper_eq_beta_i}),
expand both sides of eq.~(\ref{helper_eq_beta_i}) in $q$ and obtain the unknown coefficients $\beta_{i,j}$ by comparing
terms in each order of $q$.
We find
\bq
 \beta_{i,0} & = & 2 \arcsin u_{i,0} 
 = - i \ln v_i,
 \nonumber \\
 \beta_{i,n} & = & - 2 d_{i,n},
 \;\;\;\; n \ge 1.
\eq
Once the $q$-expansion of $\beta_i$ is known, the $q$-expansion of
\bq
 w_i & = & \sum\limits_{n=0}^\infty w_{i,n} q^n
\eq
is easily obtained from the series expansion of the exponential function.

\end{appendix}

% ----------------------------------------------
% references
\bibliography{/home/stefanw/notes/biblio}
\bibliographystyle{/home/stefanw/latex-style/h-physrev5}

\end{document}